\documentclass{jpp}
\usepackage{graphicx}

\usepackage{xcolor}
\usepackage{color}
\usepackage{graphicx}
\usepackage[utf8]{inputenc}
\usepackage[T1]{fontenc}
\usepackage{amsmath}

\usepackage{lineno} 

\shorttitle{Ponderomotive forces in non-thermal plasmas}
\shortauthor{J. Espinoza, F.A Asenjo \& P.S Moya}

\title{Ponderomotive forces in unmagnetized plasmas described by Kappa distribution functions}

\author{Joaquín Espinoza-Troni\aff{1}
  \corresp{\email{joaquinun@gmail.com}},
  Felipe A Asenjo\aff{2}\corresp{\email{felipe.asenjo@uai.cl}}
 \and Pablo S Moya\aff{1}\corresp{\email{pablo.moya@uchile.cl}}}

\affiliation{\aff{1}Departmento de Física, Facultad de Ciencias,
Universidad de Chile, Santiago, Chile \aff{2} Facultad de Ingenier\'ia y Ciencias,
Universidad Adolfo Ib\'a\~nez, Santiago, Chile.}

\begin{document}

\maketitle

\begin{abstract}
The Washimi and Karpman ponderomotive interaction due to electromagnetic waves propagation is investigated for unmagnetized plasmas described by a isotropic Kappa distribution. We performed a brief analysis of the influence of the Kappa distribution in the dispersion relations for a low temperature plasma expansion at the lowest order in which the thermal effects are appreciated. The spatial and temporal factor of the ponderomotive force is obtained as a function of the wavenumber, the spectral index $\kappa$ and the ratio between the plasma thermal velocity and the speed of light. Our results show that for unmagnetized plasmas non-thermal effects are negligible for the spatial ponderomotive force when non-relativistic thermal velocities are considered. However, for unmagnetized plasmas the temporal factor of the ponderomotive force appears only due to the presence of suprathermal particles, with a clear dependence on the $\kappa$ index. We have also analysed the role of the non-thermal effect in the induced Washimi and Karpman ponderomotive magnetization and the total power radiated associated with it. We have also shown that the slowly varying induced ponderomotive magnetic field magnitude increases as the plasma moves away from thermal equilibrium. 
\end{abstract}

\section{Introduction}

Ponderomotive forces are time averaged nonlinear forces due to high frequency electromagnetic waves in plasmas. Despite their complexity, the analysis of these forces facilities the studying of phenomena associated with electromagnetic waves and plasma interactions by simplifying their dynamics. The exact behavior of a dynamical system using the Lorentz force is very complex, and we are usually more interested in the average dynamic of the system \citep{lundin_ponderomotive_2007}. Hence, the study of ponderomotive forces it has been of great importance on the understanding of plasma phenomena in different environments. 
For example the ponderomotive force has been investigated in laser phenomena where they can lead to the appearance of self-focusing \citep{karpman_two-dimensional_1977,washimi_method_1989, rezapour_zahed_mokhtary_2018, gupta_kumar_bhardwaj_2022}. It is also known that the ponderomotive force can be deduced using a fluid formalism or by a approach using the space-time derivatives of the stress tensor \citep{washimi_ponderomotive_1976,kentwell_time-dependent_1987}. In addition the ponderomotive interaction with plasmas can act as a generator of slowly varying magnetic fields \citep{washimi_magnetic_1977}. These has been extensively investigated recently, specially for quantum plasmas, due to their importance in the magnetic field generation in laser matter interaction and in dense plasmas in astrophysical compact objects \citep{shukla_generation_2010,na_temperature_2009,jamil_karpman-washimi_2019}. \\

In relation to space physics phenomena where the plasma constantly interacts with waves the ponderomotive force plays an essential role on the physics of electromagnetic ULF (Ultra Low Frequency) waves in the terrestrial magnetosphere. Indeed considering ponderomotive forces analysis allows to model the interaction of the plasma with the waves that, in conjunction with the particles, are responsible for the transfer of energy, mass and momentum in the Earth's magnetosphere from the solar wind. Therefore the ponderomotive forces are partially responsible of phenomena like the acceleration of particles in the polar regions or the ponderomotive redistribution of plasma in the magnetosphere \citep{guglielmi_ponderomotive_2001,lundin_ponderomotive_2007,nekrasov_nonlinear_2012,nekrasov_nonlinear_2014}. Due to the contribution of these phenomena to the understanding of the dynamics of the near-Earth space environment recently it has been proposed a method to verify them experimentally \citep{guglielmi_impact_2018}. \\

On the other hand it has been observed that in near Earth space plasma the particle velocities distributions show suprathermal tails that are well described by the family of Kappa distributions \citep{lazar_kappa_2021}. These distributions depends on the spectral index $\kappa$ and can be understood as a power-law generalization from which the Maxwellian distribution is recovered as a limit case when $\kappa$ tend to infinity. However, a detailed analysis of the effect of the dispersion relation in the ponderomotive interaction of electromagnetic waves with non-thermal plasmas has not been investigated yet. As explained above, in the near-Earth space environment it has been observed that plasmas are mainly described by Kappa distributions  \citep{Espinoza_2018,Moya_2021,Eyelade_2021}. Therefore, to better understand the dynamics of our space environment, the impact of the non-thermal effect of the plasma in its interaction with the waves that propagate in the Earth's atmosphere must be evaluated. \\

The ponderomotive force not only depends in the spatial and temporal variations of the wave amplitude but also strongly depends in the properties of the medium and its interaction with the electromagnetic waves described by its dielectric tensor, therefore this nonlinear effect principally stands on the distribution of the particles that describes the plasma. Hence, the non-thermal effects of the plasma described by the Kappa distribution can significantly affect the behavior of the ponderomotive force and thus the phenomena associated with it.	This can lead to great impact in space physics where as we said above various phenomena related with ponderomotive forces occurs in plasmas described by Kappa distributions. In fact, as we know from the work of \citep{kim_nonthermal_2009} the non-thermal effect plays a significant role in the temporal term of the ponderomotive force for electrostatic waves. Where is it show that in consequence the Washimi and Karpman induced magnetization its enhanced because of the non-thermal effects. However to the best of our knowledge, the impact of the non-thermal effects due to Kappa distribution in the poderomotive force due to the electromagnetic waves interaction with plasmas has not been studied. The purpose of this work is to study the influence of the Kappa distribution in the Washimi and Karpman ponderomotive force for waves in unmagnetized plasmas and for all of its modes of propagation. This research can be useful to generalize the results given by \citep{hora_1969,gupta_kumar_2021}. We are going to include this effects using the dielectric tensor for Kappa distributions in the low temperature approximation and we are going to compare the magnitude of the term that accompanies the variation of the wave amplitude for the Kappa and Maxwellian distributions. Hence, this work can be also useful for understand the implications of the plasma temperature in the ponderomotive force interaction. \\

This article is organized as follows: in section 2, we include the non-thermal effects in the dielectric tensor and we expand it for low temperatures. We also give a brief analysis of the influence of the kappa parameter in the transverse and longitudinal dispersion relations. Then, in section 3 we include the dielectric tensor in the spatial and temporal terms of the ponderomotive force and we deduce its expressions. Later we analyze for each term of the ponderomotive force and wave mode the influence of the non-thermal effects and we compare it with the Maxwellian case. In section 4 we use this analysis to study the Washimi and Karpman ponderomotive induced magnetic field for electromagnetic waves. Finally, in section 5 we summarize the main conclusions of this work.

\section{Dispersion relation for thermal and non-thermal plasmas with low temperature}

In this investigation we are going to use a isotropic Kappa distribution, with its corresponding dispersion function \citep{summers_modified_1991}. In this work we are going to consider the simplest form of the Kappa distribution \citep{Lazar2016}
\begin{equation}
    f_{\kappa s}(\mathbf{v}) = \frac{n_s}{\pi^{3/2}\alpha_s^3 \kappa^{3/2}}\frac{\Gamma(\kappa+1)}{\Gamma(\kappa-1/2)}\left(1 + \frac{v^2}{\kappa \alpha_s^2}\right)^{-(\kappa + 1)} 
\label{eq:DistribucionKappa}
\end{equation}
Here, $f_{\kappa s}$ is the Kappa distribution for the species $s$, $n_s$ is their number density, $\alpha_s\{= \sqrt{2k_B T_s /m_s}\}$ is their thermal velocity, $k_B$ is the Boltzman constant, $m_s$ is their mass, $T_s$ is their temperature and $\Gamma$ is the Gamma function. \\

In this section we are going to analyze the dispersion relation for low temperature plasmas described by a Kappa distribution (\ref{eq:DistribucionKappa}) and without a background magnetic field. A detailed deduction of the dielectric tensor and the dispersion relation for non-relativistic magnetized plasma modelled by isotropic Kappa distributions can be found in \citet{mace_dielectric_1996}. The unmagnetized dispersion relation for transverse and longitudinal modes with respect to the direction of wave propagation are given respectively by
\begin{equation}
    \frac{c^2k^2}{\omega^2} = 1 + \sum_s \frac{\omega_{ps}^2}{\omega k \alpha_s} Z_{\kappa M} \left(\frac{\omega}{k\alpha_s}\right)
\end{equation}

\begin{equation}
    1 - \sum_s \left(\frac{\omega_{ps}^2}{k^2\alpha_s^2}\right)Z'_{\kappa M}\left(\frac{\omega}{k\alpha_s}\right) = 0
\end{equation}
where $\omega$ is the frequency, $k$ is the wave number, $c$ is the velocity of light,  $Z_{\kappa M}$ is the generalized plasma dispersion function \citep{hellberg_generalized_2002}, and $\omega_{ps}$ is the plasma frequency for the species $s$. \\

To consider low temperature plasmas ($\omega/k\alpha_s \gg 1$) we  expand the generalized plasma dispersion function for large arguments at the lowest order in which the effect of the kappa parameter appears, so that the temperature is included in the dielectric tensor. In this way, we expand in $k\alpha_s/\omega$ at second order for the transverse component of the dielectric tensor and at third order for the longitudinal component of the dielectric tensor.  We do not consider the imaginary terms because we are not interested in the damping characteristics of the wave. Accordingly, we use the following expansion \citep{hellberg_generalized_2002}
\begin{equation}
    Z_\kappa^*(\zeta) \approx -\frac{1}{\zeta}\left(1 + \frac{\kappa}{2\kappa-3}\frac{1}{\zeta^2} + \frac{3\kappa^2}{(2\kappa-5)(2\kappa-3)}\frac{1}{\zeta^4} + 15\frac{\kappa^3}{(2\kappa-7)(2\kappa-5)(2\kappa-3)}\frac{1}{\zeta^6} + \cdots\right)
\end{equation}

By considering only the electron species, we have the approximated dispersion relation for transverse and longitudinal modes respectively as follows
\begin{equation}
   \varepsilon_\perp \approx 1 - \frac{1}{\bar{\omega}^2}\left(1+ \frac{1}{2}\left(\frac{\alpha_e^2}{c^2}\right)\frac{\kappa}{\kappa-3/2} \frac{\bar{k}^2}{\bar{\omega}^2}\right)
   \label{eq:RDtransversal}
\end{equation}

\begin{equation}
    \varepsilon_\parallel \approx 1 - \frac{1}{\bar{\omega}^2}\left(1 + \frac{3}{2}\left(\frac{\alpha_e^2}{c^2}\right)\frac{\kappa}{\kappa - 3/2} \frac{\bar{k}^2}{\bar{\omega}^2}\right) 
    \label{eq:RDlongitudinal}
\end{equation}
where $\varepsilon_\perp$ and $\varepsilon_\parallel$ are the dielectric tensor components for the transverse and longitudinal modes respectively, and $\bar{\omega} = \omega/\omega_{pe}$ and $\bar{k} = kc/\omega_{pe}$ are normalized frequency and wavenumbers.  Notice in equation (\ref{eq:RDlongitudinal}) that we  recover the result obtained by Ziebell \textit{et al} for Langmuir waves with a plasma distributed by a isotropic Kappa
distribution of type I \citep{ziebell_dispersion_2017}. If we include the dispersion relation in the dielectric components we have that $\epsilon_\perp = \bar{k}^2/\bar{\omega}^2$ and $\varepsilon_\parallel = 0$. So if we consider that $\alpha_e \ll c $ we can express the transversal dielectric tensor component as function of the frequency as

\begin{equation}
  \varepsilon_\perp \approx \left(1 + \frac{1}{\bar{\omega}^2}\right)\left(1-\frac{1}{2}\left(\frac{\alpha_e}{c}\right)^2\frac{\kappa}{\kappa-3/2}\frac{1}{\bar{\omega}^2}\right) 
   \label{eq:RDtransversalFrequency}
\end{equation}

We can also calculate analytically the solutions for the dispersion relation for the transverse and longitudinal modes which are respectively given by

\begin{equation}
    \bar{\omega}^2_{\kappa\perp} = \frac{1}{2}\left(\bar{k}^2 + 1\right) + \frac{1}{2}\sqrt{\left(\bar{k}^2 + 1\right)^2 + 2\left(\frac{\alpha_e^2}{c^2}\right)\frac{\kappa}{\kappa-3/2}\bar{k}^2}
    \label{eq:RDtransversal_solucion}
\end{equation}
 
\begin{equation}
    \bar{\omega}^2_{\kappa\parallel} = \frac{1}{2} + \frac{1}{2}\sqrt{1 + 6\left(\frac{\alpha_e^2}{c^2}\right)\frac{\kappa}{\kappa-3/2}\bar{k}^2}
    \label{eq:RDlongitudinal_solucion}
\end{equation}
where we have included the subscript $\kappa$ to make explicit its dependence on kappa, and the subscript $\perp$ ($\parallel$) to indicate the solution for transverse (longitudinal) waves. In  dispersion relations (\ref{eq:RDtransversal_solucion}) and (\ref{eq:RDlongitudinal_solucion}) it can be seen that unlike the electromagnetic waves the electrostatic waves propagation occurs only due to the thermal effect that is contained in the thermal velocity. It is therefore to be expected that the effect of the spectral index $\kappa$ will be more notorious for the longitudinal modes than for the transverse modes. In both cases, when $\kappa\rightarrow\infty$, we recover classical Maxwellian results that we are going to denote with the subscript $\mathsf{M}$   in the frequency and in the ponderomotive force (see below). \\

\begin{figure}
\centering
\includegraphics[height=2.3in]{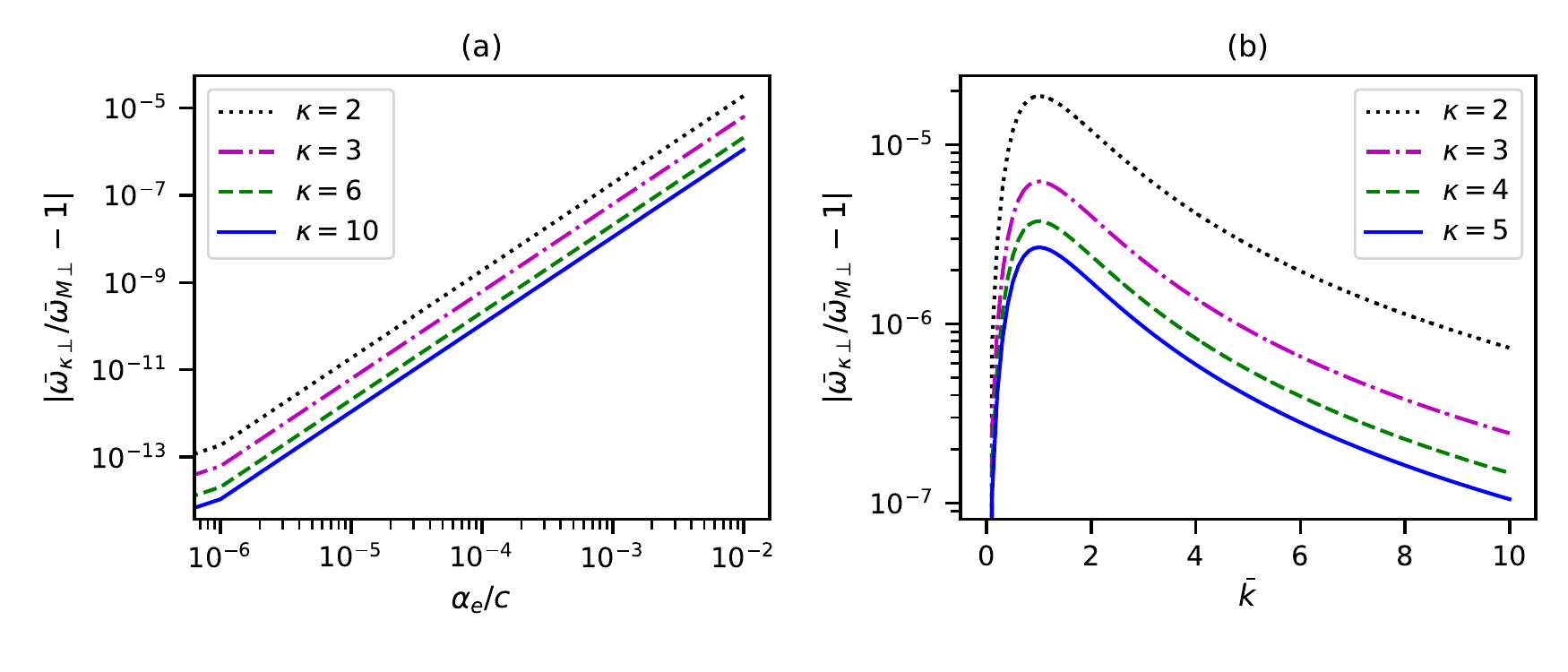}
\caption{(a) Relative difference of the frequencies $|\bar\omega_{\kappa\perp}/\bar\omega_{\textsf{M}\perp}-1|$ for the transverse dispersion relation (\ref{eq:RDtransversal_solucion}) as function of $\alpha_e/c$ with $\bar{k} = 1$ at logarithmic scale. (b) Relative difference of the frequencies $|\bar\omega_{\kappa\perp}/\omega_{\textsf{M}\perp}-1|$ for the transverse dispersion relation (\ref{eq:RDtransversal_solucion}) as function of $\bar{k}$ with $\alpha_e/c = 10^{-2}$ at logarithmic scale. }
\label{fig:RDtransversal}
\end{figure}

\begin{figure}
  \centering
  \includegraphics[width=0.6\linewidth]{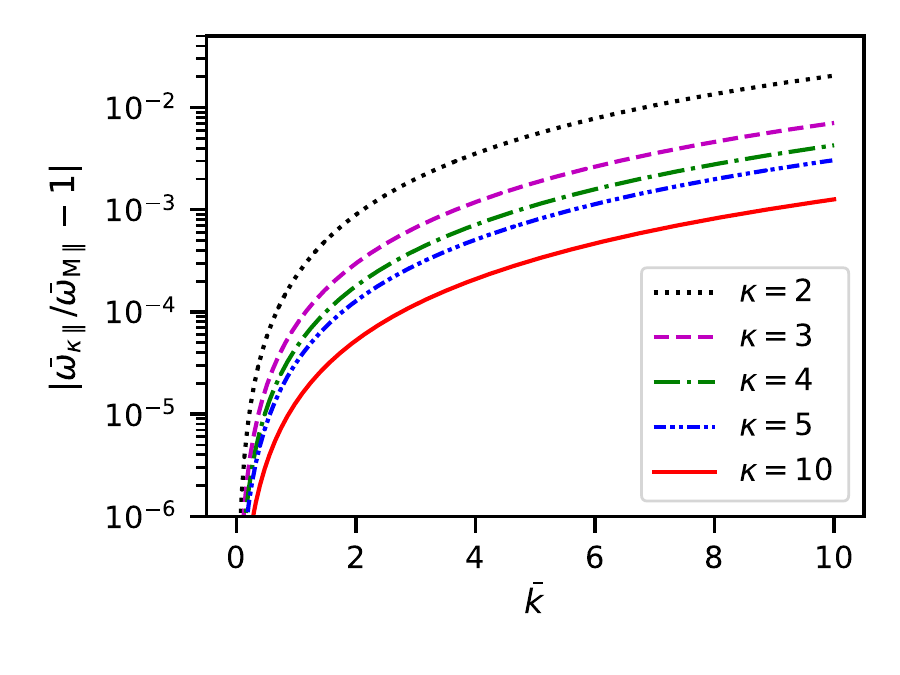}
  \caption{Relative difference of the frequencies $|\bar\omega_{\kappa\parallel}/\bar\omega_{\textsf{M}\parallel}-1|$ for longitudinal waves (\ref{eq:RDlongitudinal_solucion}) as function of $\bar{k}$ with $\alpha_e/c = 10^{-2}$ at logarithmic scale.}
  \label{fig:RDlongitudinal}
\end{figure}

Fig. \ref{fig:RDtransversal}.(a) represents the relative difference of the frequencies for the transverse waves comparing the Kappa and Maxwellian distribution $|\bar{\omega}_{\kappa\perp}/\bar{\omega}_{\textsf{M}\perp} - 1|$ as a function of the velocity ratio $\alpha_e/c$ for various values of $\kappa$. As we see in this figure the scaled frequency in the transverse dispersion relation varies with the kappa parameter at the order of $\sim 10^{-5}$ or less relative with the Maxwellian case for non-relativistic thermal velocity values ($\alpha_e/c < 10^{-2}$), unlike the longitudinal modes that varies at the order of $\sim 10^{-2}$ with respect to the Maxwellian case like it can be seen in the Fig. \ref{fig:RDlongitudinal} which represents the solution for the dispersion relation for longitudinal waves as function of the scaled frequency $\bar{k}$ for various values of $\kappa$. \\

It can be seen in equation (\ref{eq:RDtransversal}) that the term that contains the kappa parameter is multiplied by $\alpha_e^2/c^2$ so for the transverse waves in non-relativistic plasmas the thermal effects are not significant unless we have a plasma very far from thermal equilibrium with $\kappa \sim 3/2$. \\

Figure \ref{fig:RDtransversal}.b represents the relative difference of the frequencies for the transverse waves comparing the Kappa and Maxwellian distribution $|\bar{\omega}_{\kappa\perp}/\bar{\omega}_{\textsf{M}\perp} - 1|$ as a function of the scaled wavenumber $\bar{k}$ for various values of $\kappa$. As we can see in this figure for lower wavenumbers different from zero the relative difference between the scaled frequency for the Kappa distribution and Maxwellian distributions tends to a maximum which we can demonstrate analytically that occurs at $\bar k=1$. And for non-relativistic plasmas with $\alpha_e/c \sim 10^{-2}$ and $\kappa < 6$ this maximum is of the order of $10^{-5}$. We can note that the maximum occurs when the wavelength $\lambda$ is equal to the inertial length $c(2\pi/\omega_{pe})$ regardless of the thermal velocity or the kappa parameter.

\section{Ponderomotive Force}

The ponderomotive force is a non-linear phenomena induced by the interaction of a high-frequency field in a slow time scale motion with the plasma \citep{kentwell_time-dependent_1987}. The purpose of this work is to study how the non-thermal effect described by the Kappa distribution impacts in the non-linear slow time scale interaction of unmagnetized plasmas with high frequency electromagnetic fields. In this discussion we are going to use the Washimi and Karpman ponderomotive force formalism \citep{karpman_ponderomotive_1982} and the previous results  for the dispersion relations for Lorentzian cold plasmas.  \\

In the absence of a background magnetic field the total Washimi-Karpman ponderomotive force  $\mathbf{f}_{\text{WK}} = \mathbf{f}_{(s)} + \mathbf{f}_{(t)}$, due to the electromagnetic field  $\bar{\mathbf{E}}(\mathbf{r},t)\{=(1/2)[\mathbf{E}(\mathbf{r},t)e^{i(\mathbf{k}\cdot \mathbf{r} - \omega t)} + \mathbf{E}^*(\mathbf{r},t)e^{-i(\mathbf{k}\cdot \mathbf{r} - \omega t)}]\}$,  is described by a force $\mathbf{f}_{(s)}$ associated to the spatial variation of the electric field amplitude $|\mathbf{E}|$ and a force $\mathbf{f}_{(t)}$ associated with its temporal variations. 
For a isotropic medium we can express the spatial-variation part of the ponderomotive force as follows
\begin{equation}
    \mathbf{f}_{(s)} = \frac{1}{8\pi}(\varepsilon - 1)\nabla |\mathbf{E}|^2
    \label{eq:FPespacial}
\end{equation}
where   $\varepsilon$ is the component of the dielectric tensor for the transverse or longitudinal mode, as appropriated. If we also suppose that the magnitude of the electric field varies slowly in our time and space scales and we discard second time derivatives ($(\partial |\mathbf{E}| /\partial x)(\partial |\mathbf{E}|/\partial t) \ll 1$ and $\partial |\mathbf{E}|^2/\partial x\partial t \ll 1$)  it can be deduced that the temporal-variation part of the ponderomotive force becomes \citep{washimi_ponderomotive_1976}
\begin{equation}
    \mathbf{f}_{(t)} = \frac{\mathbf{k}}{16\pi\omega^2} \frac{\partial \omega^2 (\varepsilon - 1)}{\partial \omega}\frac{\partial |\mathbf{E}|^2}{\partial t}
    \label{eq:FPtemporal}
\end{equation}

Notice that to compute the partial derivative of $\varepsilon$ in the temporal term of the ponderomotive force we have to use equation (\ref{eq:RDtransversal}) instead of (\ref{eq:RDtransversalFrequency}). Now let us analyze the factors $f_{(s)}\{=(1/8\pi)(\varepsilon - 1)\}$ and $f_{(t)}\{= (k/16\pi \omega^2)[\partial \omega^2(\varepsilon - 1)/\partial \omega] \}$ that accompany the spatial and temporal variations of the magnitude of the electric field in the ponderomotive force for unmagnetized plasmas of electrons species described by Kappa distributions. \\

\subsection{Spatial ponderomotive force factor for longitudinal waves}

In this case, 
   the medium has no effect in the ponderomotive force, since   $\varepsilon_\parallel = 0$. \\

\subsection{Spatial ponderomotive force factor for transverse waves} 

If we include the dispersion relation in the ponderomotive force, using the equations (\ref{eq:RDtransversalFrequency}) and (\ref{eq:FPespacial}) we can deduce that the factor $f_{(s)}^\kappa$ that accompany the spatial variation in the ponderomotive force for the Kappa distribution is given by:

\begin{equation}
    f_{(s)}^{\kappa\perp} = -\frac{1}{8\pi \bar{\omega}_{\kappa\perp}^2} - \frac{1}{16\pi}\left(\frac{\alpha_e^2}{c^2}\right)\left(\frac{\kappa}{\kappa-3/2}\right)\frac{1}{\bar{\omega}_{\kappa\perp}^2}\left(1-\frac{1}{\bar{\omega}_{\kappa\perp}^2}\right)
\label{eq:FPespacial_solucion}
\end{equation}
where the superindex $\kappa$ indicates its dependence in the kappa parameter,   $\perp$ indicates that we are considering the ponderomotive force for transverse waves, and $\bar{\omega}_{\kappa,\perp}$ corresponds to the solution for the dispersion relation for the parameter $\kappa$ for the transverse waves  (\ref{eq:RDtransversal_solucion}). Also, for the Maxwellian ponderomotive force ($\kappa \rightarrow \infty$) we will denote with the superindex $\mathsf{M}$.

\begin{figure}
\centering
\includegraphics[width=\linewidth]{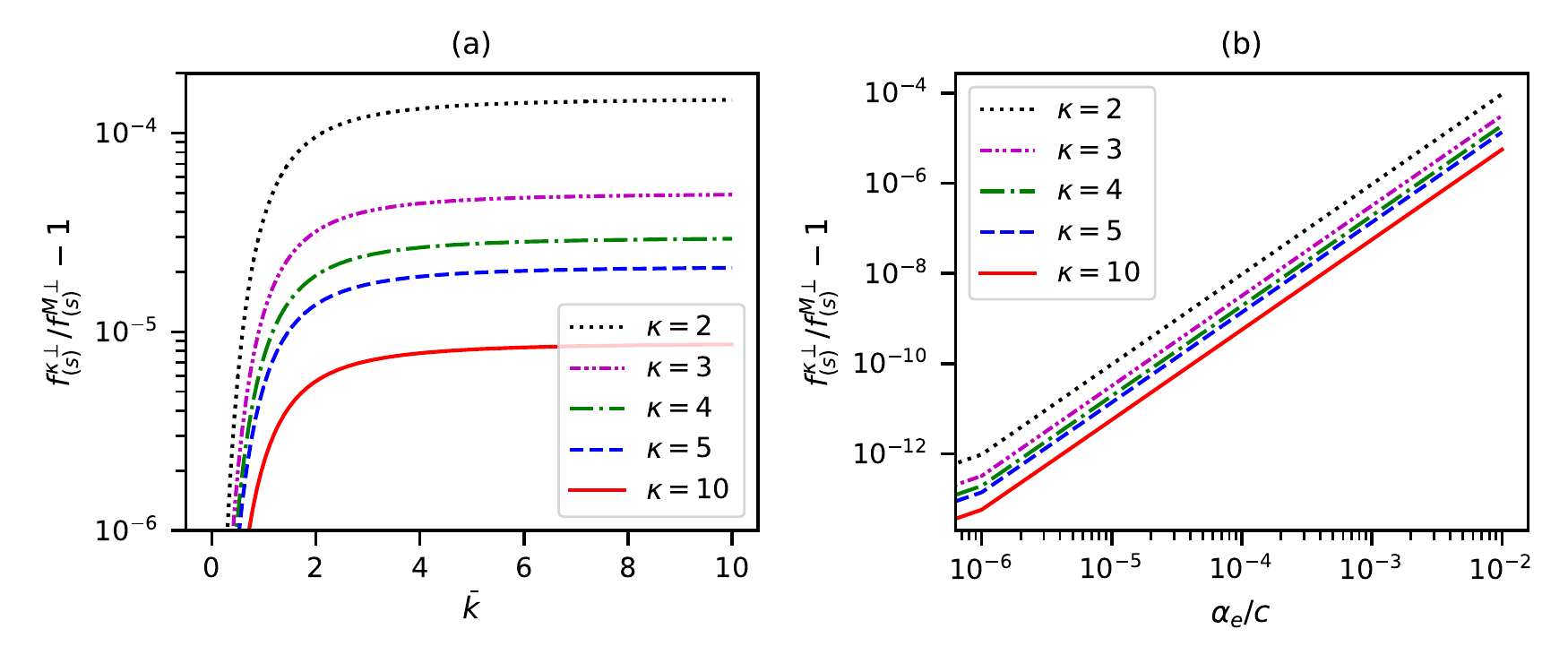}
\caption{Relative difference of the Kappa ponderomotive spatial factor and the Maxwellian ponderomotive spatial factor for transverse waves $f^\kappa_{(s)}/f^{\textsf{M}}_{(s)}-1$ at logarithmic scale (a) as function of $\bar{k}$ with $\alpha_e/c = 0.01$ equation (\ref{eq:FPespacial_solucion}).  (b) as function of $\bar{c}$ with $\bar{k} = 2$.}
\label{fig:FPespacial}
\end{figure}

Figure 3.(a) shows the relative difference of the Kappa and Maxwellian ponderomotive spatial factor for transverse waves $f_{(s)}^{\kappa\perp}/f_{(s)}^{\textsf{M}\perp} - 1$ as function of the scaled wave number. As it is seen, the magnitude of the ponderomotive force for the Kappa distributions is greater compared to the Maxwellian distribution case, and as the scaled wavenumber decreases both magnitudes are equal so the non-thermal effect is canceled. We can also see that for $\bar{k} \rightarrow \infty$ the rate between the Kappa and Maxwellian ponderomotive spatial factors tends to an asymptotic value.  Figure 3.b represents the relative difference of the Kappa and Maxwellian ponderomotive spatial factor for transverse waves as a function of  $\alpha_e/c$. As we can see in this figures the non-thermal effect on the ponderomotive force increases with the thermal velocity scaled with the velocity of light and in this case that we are dealing with non-relativistic thermal velocities the effect of the spectral index $\kappa$ is negligible, having a relative difference between the ponderomotive forces of an order less than or equal to $\sim 10^{-4}$ for $\kappa > 2$ and $\alpha_e/c < 10^{-2}$. This result is reasonable considering the previous discussion of the dispersion relation where it was concluded that the thermal effect is not appreciable for non-relativistic thermal velocities. \\

In the limit of large wavenumbers (which serves as an upper bound) we have that the ratio for the Kappa and Maxwellian ponderomotive spatial terms tends to $\left[1 + \frac{1}{2}(\alpha_e^2/c^2)\left(\frac{\kappa}{\kappa - 3/2}\right)\right]/[1 + \frac{1}{2}\left(\alpha_e^2/c^2\right)] > 1$  where it can be seen clearly that when $\alpha_e^2/c^2 \ll 1$ the effect of the kappa parameter is canceled. Also we can calculate that for $\kappa > 2$ and $\alpha_e/c = 10^{-2}$ the spatial term of the ponderomotive force for the Lorentzian plasmas is bigger than for the Maxwellian plasmas for less than $0.015\%$. \\
 
\subsection{Temporal ponderomotive force factor for transverse waves}

In this section we are going to analyze the non-thermal effect in the temporal factor of the ponderomotive force for the transverse waves. Using the Washimi and Karpman expression of the temporal factor (\ref{eq:FPtemporal}) and including the dispersion relation (\ref{eq:RDtransversal}) we can find the expression for the temporal term of the ponderomotive force for the transverse waves
\begin{equation}
    f_{(t)}^{\kappa\perp} = \frac{1}{16\pi c}\left(\frac{\alpha_e^2}{c^2}\right) \left(\frac{\kappa}{\kappa-3/2}\right)\frac{\bar{k}^3}{\bar{\omega}_{\kappa\perp}^5}
    \label{eq:FPtemporal_transversal}
\end{equation}

\begin{figure}
    \centering
    \includegraphics[width=0.6\linewidth]{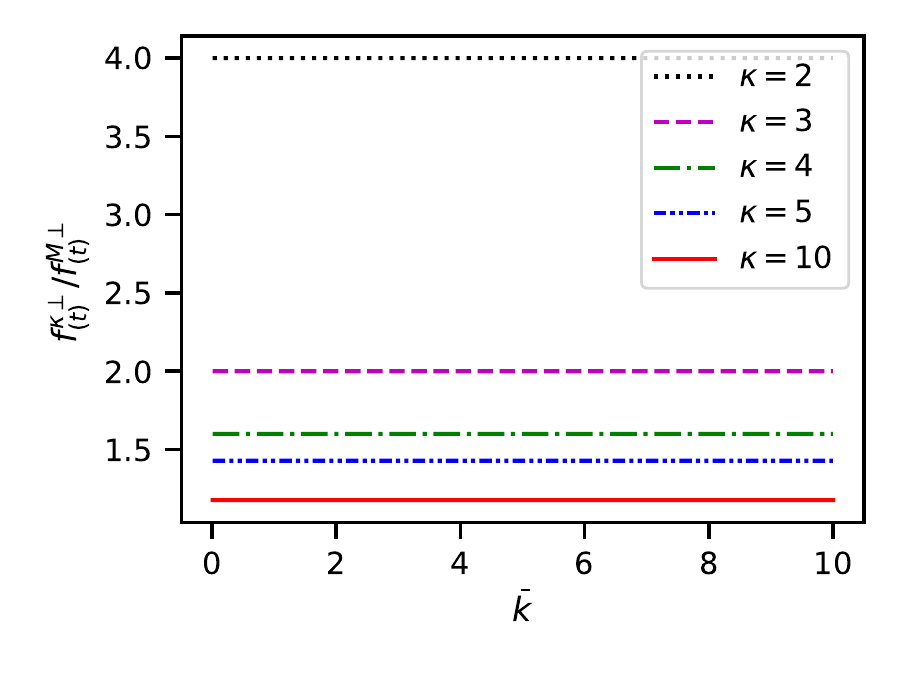}
    \caption{Relative difference of the Kappa ponderomotive temporal factor and the Maxwellian ponderomotive temporal factor for transverse waves $f^{\kappa\perp}_{(t)}/f^{\textsf{M}\perp}_{(t)}$ at logarithmic scale as function of $\bar{k}$ with $\alpha_e/c = 10^{-2}$ equation (\ref{eq:FPtemporal_transversal}). }
\label{fig:FPespacial}
    \label{fig:FPtemporal_transversal}
\end{figure}

It follows from the previous equation that the temporal factor of the ponderomotive force is nonzero only when we consider the effects of the temperature in the dielectric tensor \citep{washimi_ponderomotive_1976} . Hence, the temporal term of the Washimi and Karpman ponderomotive force for unmagnetized plasmas is a thermal effect. Figure \ref{fig:FPtemporal_transversal} shows the relative difference of the Kappa and Maxwellian ponderomotive temporal factor for transverse waves as function of $\bar{k}$. As we can see in this figure the non-thermal effect on the temporal term of the ponderomotive force it is very significant and it remains almost constant with respect to the scaled wavenumber. We can also deduce in the expression (\ref{eq:FPtemporal_transversal}) that for any value of $\kappa$ the ratio between the forces does not vary significantly with thermal velocity. Also the effect of the kappa parameter is given mainly by the term $\kappa/(\kappa-3/2)$ that accompanies the other dispersion relation dependent terms by reason of the previous section where we know that if $\alpha_e/c \ll 1$ then  $\bar{\omega}_{\kappa\perp}/\bar{\omega}_{\textsf{M}\perp} \approx 1$. So for non-relativistic velocities we can use the following approximation $ f^{\kappa\perp}_{(t)}/f^{\mathsf{M}\perp}_{(t)} \approx \kappa/(\kappa-3/2) > 1$. \\

At first glance we can see that this expression does not depend on the thermal velocity and only depends on the parameter $\kappa$. We can calculate that for $\kappa < 6$ the temporal factor of the ponderomotive force for the Lorentzian plasmas is bigger than for the Maxwellian plasmas for at least a 33\%. Hence, due to the thermal character of the ponderomotive temporal factor for unmagnetized plasmas the non-thermal effect is strongly noticeable in the temporal ponderomotive term for transverse waves. \\

\subsection{Temporal ponderomotive force factor for longitudinal waves}

In this section we are going to analyze the non-thermal effect in the temporal factor of the ponderomotive force for longitudinal waves. Using the Washimi and Karpman expression of the temporal factor (\ref{eq:FPtemporal}) and including the dispersion relation (\ref{eq:RDlongitudinal}) we can find the expression for the temporal term of the ponderomotive force for longitudinal waves:

\begin{equation}
\begin{split}
    f^{\kappa\parallel}_{(t)} &= \frac{3}{16\pi c}\left(\frac{\alpha_e^2}{c^2}\right) \left(\frac{\kappa}{\kappa-3/2}\right)\frac{\bar{k}^3}{\bar{\omega}_{\kappa\parallel}^5} \\
    \label{eq:FPtemporal_longitudinal}
\end{split}
\end{equation}
where $\parallel$ indicates that we are talking for the ponderomotive force for longitudinal waves, while $\bar{\omega}_{\kappa,\parallel}$ corresponds to the solution for the dispersion relation for the parameter $\kappa$ for the longitudinal waves  (\ref{eq:RDlongitudinal_solucion}). \\

\begin{figure}
    \centering
    \includegraphics[width=0.6\linewidth]{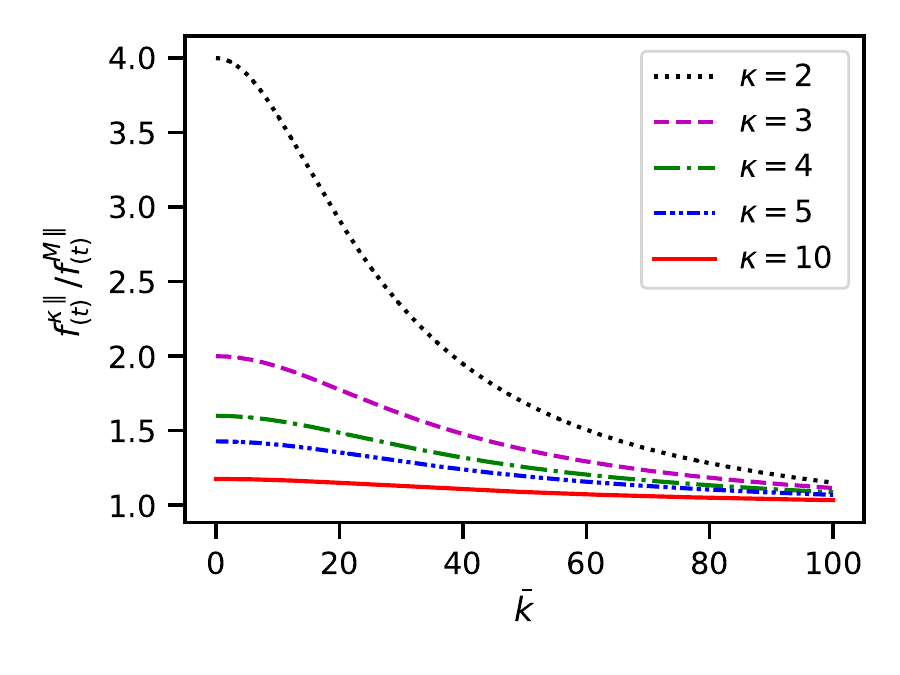}
    \caption{Relative difference of the Kappa ponderomotive temporal factor and the Maxwellian ponderomotive temporal factor for longitudinal waves $f^{\kappa\parallel}_{(t)}/f^{\textsf{M}\parallel}_{(t)}$  at logarithmic scale as function of $\bar{k}$ with $\alpha_e/c = 0.01$ equation (\ref{eq:FPtemporal_longitudinal}). }
\label{fig:FPespacial}
    \label{fig:FPtemporal_longitudinal}
\end{figure}

We can see that this equation is very similar to the transverse waves ponderomotive temporal factor. Figure \ref{fig:FPtemporal_longitudinal} represents the relative difference of the Kappa and Maxwellian ponderomotive temporal factors as function of the scaled wave number $\bar{k}$. Hence, including the longitudinal dispersion relation in the ponderomotive temporal factor it can be seen in this figure that for wavenumbers less than some value the temporal factor is bigger for Lorentzian plasmas than Maxwellian plasmas and for low wavenumbers its ratio tends to behave as $\kappa/(\kappa-3/2) > 1$ that is the same behaviour as the transversal case, nevertheless because the longitudinal waves are more affected by the dispersion relation we can see that the non-thermal effect varies with the wavenumber and tend to decrease as the wavenumber increases. For bigger wavenumbers the temporal factor is lower for Lorentzian plasmas than Maxwellian plasmas and tends to behave as $ \left((\kappa-3/2)/\kappa\right)^{1/4} < 1$ but we are not interested in such big wavenumbers because in the fluid scale that we are looking they don't have relevance. Hence, we can conclude that as well as for the transverse waves,  the non-thermal effect in the temporal ponderomotive term for longitudinal waves  is also strongly noticeable. \\

\section{Washimi and Karpman ponderomotive magnetization}

According to the work of Washimi and Watanabe \citep{washimi_magnetic_1977} a slowly varying magnetic field $\mathbf{B}_0$ is generated by the ponderomotive force of the electromagnetic wave that is slowly varying in time. From the balance of the ponderomotive force with the slowly varying electromagnetic field in the electron-fluid equation of motion it follows that the induced magnetic field is given by:

\begin{equation}
    \mathbf{B}_0(\mathbf{r},t) = -\frac{c}{16\pi n_e e \omega^2}\frac{\partial [\omega^2(\varepsilon - 1)]}{\partial \omega}\nabla\times (\mathbf{k}|\mathbf{E}|^2)
    \label{eq:WKponderomotiveMagneticField}
\end{equation}
where $e$ is the electron charge.  Kim and Jung have calculated the Washimi and Karpman ponderomotive magnetic field for the electrostatic case \citep{kim_nonthermal_2009}. Hence, in this section we are going to perform this calculation for the electromagnetic case using the dielectric tensor approximation for transverse waves in low temperature plasmas expanding by one order of magnitude more than in (\ref{eq:RDtransversal}) so that $\alpha_e/c$ appears in the ratio of the magnetizations and we can see its effect (see below)

\begin{equation}
    \varepsilon(\bar{k},\bar{\omega}) \approx 1 - \frac{1}{\bar{\omega}^2}\left[1 + \frac{1}{2}\left(\frac{\alpha_e^2}{c^2}\right)\left(\frac{\kappa}{\kappa-3/2}\right)\frac{\bar{k}^2}{\bar{\omega}^2} + \frac{3}{4}\left(\frac{\alpha_e^4}{c^4}\right)\frac{\kappa^2}{(\kappa-5/2)(\kappa-3/2)}\frac{\bar{k}^4}{\bar{\omega}^4}\right]
    \label{eq:ElectromagneticDielectricTensor}
\end{equation}
 
Inserting the equation (\ref{eq:ElectromagneticDielectricTensor}) in (\ref{eq:WKponderomotiveMagneticField}) the magnitude $B_\kappa$ of the ponderomotive magnetic field for Lorentzian plasmas is obtained as:

\begin{equation}
    B_\kappa \approx  \frac{c}{8\pi n_e e}\frac{1}{\omega_{pe}}\left[\frac{1}{2}\left(\frac{\alpha_e^2}{c^2}\right)\left(\frac{\kappa}{\kappa-3/2}\right)\frac{\bar{k}^2}{\bar{\omega}^5} + \frac{3}{2}\left(\frac{\alpha_e^4}{c^4}\right)\frac{\kappa^2}{(\kappa-5/2)(\kappa-3/2)}\frac{\bar{k}^4}{\bar{\omega}^7}\right]\frac{k|E|^2}{L}
\end{equation}
where $L$ is the scale length of the intensity of the field. Using this we can calculate the scaled electron cyclotron frequency $\bar{\omega}_{ce} = \omega_{ce}/\omega_{pe}$ generated by the induced magnetic field:

\begin{equation}
\begin{split}
    \bar{\omega}_{ce} &= \frac{1}{16\pi}\left(\frac{\alpha_e^4}{c^4}\right)\left(\frac{\kappa}{\kappa-3/2}\right)\frac{\bar{k}^4}{\bar{\omega}^5}\left[1+ 3\left(\frac{\alpha_e^2}{c^2}\right)\left(\frac{\kappa}{\kappa-5/2}\right)\frac{\bar{k}^2}{\bar{\omega}^2}\right]\frac{\lambda}{L}\left(\frac{e|\mathbf{E}|}{m_e\lambda_{De}\omega_{pe}^2}\right)^2 \\
    &= M_p(\kappa,\bar{k},\bar{\omega})\frac{\lambda}{L}\left(\frac{u_e}{\lambda_{De}\omega_{pe}}\right)^2
\end{split}
\end{equation}
where we have defined the Karpman–Washimi ponderomotive magnetization $M_p(\kappa,\bar{k},\bar{\omega})$ as in Kim and Jung's work \citep{kim_nonthermal_2009}:

\begin{equation}
M_p(\kappa,\bar{k},\bar{\omega}) = \frac{1}{16\pi}\left(\frac{\alpha_e^4}{c^4}\right)\left(\frac{\kappa}{\kappa-3/2}\right)\frac{\bar{k}^4}{\bar{\omega}^5}\left[1+ 3\left(\frac{\alpha_e^2}{c^2}\right)\left(\frac{\kappa}{\kappa-5/2}\right)\frac{\bar{k}^2}{\bar{\omega}^2}\right]
\label{eq:WKmagnetization}
\end{equation}
with $\lambda$  the wavelength of the wave, $\lambda_{De}(= \alpha_e/\sqrt{2}\omega_{pe})$ is the Debye length of the electrons and $u_e = e|\mathbf{E}|/m_e\omega_{pe}$ is the quiver velocity \citep{kourakis_magnetization_2006}. Then it can be defined analogously in Kim and Jung's work a non-thermal effect $F_{\text{NT}}$ as the ratio of the magnetization for the Kappa distribution and for the Maxwellian distribution ($\kappa \rightarrow \infty$) 

\begin{equation}
F_{NT} = \frac{M_p(\kappa,\bar{k},\bar{\omega})}{M_p(\kappa \rightarrow \infty,\bar{k},\bar{\omega})} = \left(\frac{\kappa}{\kappa-3/2}\right)\left[\frac{1+3(\frac{\alpha_e^2}{c^2})(\frac{\overline{k}^2}{\overline{\omega}^2})\frac{\kappa}{(\kappa-5/2)}}{1+3(\frac{\alpha_e^2}{c^2})\frac{\overline{k}^2}{\overline{\omega}^2}}\right]
\label{eq:NonthermalEffect}
\end{equation}
 
 In the equation (\ref{eq:NonthermalEffect}) it can be seen that the effect of the kappa distribution is mostly represented by the term $\kappa/(\kappa-3/2)$ for non-relativistic plasmas where $\alpha_e/c \ll 1$ which means that the non-thermal effect is enhanced for lower values of $\kappa$ as it is expected. \\

\begin{figure}
\centering
  \centering
  \includegraphics[width=0.6\linewidth]{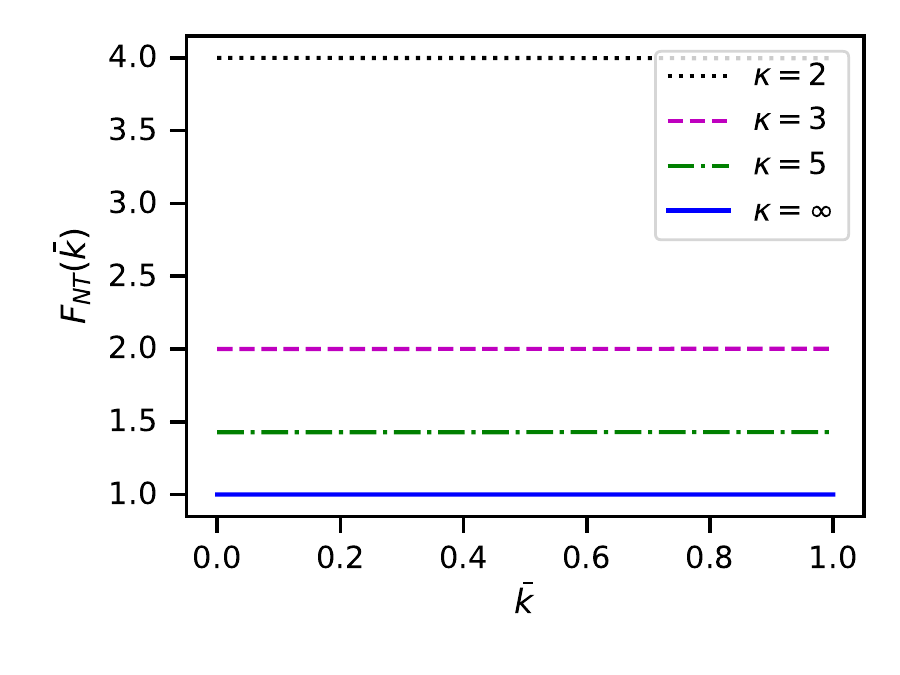}
\caption{Non-thermal effect $F_{NT}(\bar{k},\kappa)$ equation (\ref{eq:NonthermalEffect}) as a function of $\bar{k}$ for different values of $\kappa$ for $\alpha_e/c = 0.01$.}
\label{fig:PonderomotiveMagnetization}
\end{figure}

Figure \ref{fig:PonderomotiveMagnetization} represents the non-thermal effect $F_{NT}(\bar{k},\kappa)$ (\ref{eq:NonthermalEffect}) as a function of the scaled frequency $\bar{k}$ for different values of $\kappa$. It is found in this figure that the non-thermal effect decreases with an increase in the spectral index $\kappa$. Therefore, the non-thermal effect of the Kappa distributions enhances the induced magnetization due to the electromagnetic ponderomotive interactions in unmagnetized plasmas. We can also see in this figure that the effect of the Kappa distributions in the ponderomotive magnetization does not depend in the wavenumber, because as we said previously for non-relativistic plasmas this effect is mainly subject to the parameter $\kappa/(\kappa-3/2)$. It is important to note also that the non-thermal effect changes significantly the ponderomotive magnetization with a relative difference with the Maxwellian case of at least $30\%$ for $\kappa < 6$ and for non-relativistic plasmas even if $\alpha_e \rightarrow 0$.  In the non-relativistic limit we can calculate the total radiated power $P$ average in one period produced by the gyromotion of the charges by the induced Washimi and Karpman magnetic field \citep{na_temperature_2009} using the Larmor formula \citep{jackson_classical_1975} $P = \frac{2}{3}\frac{e^2r_L^2}{c^2}\left(\frac{\lambda}{L}\right)^4\left(\frac{u_e^2}{\lambda_{De}^2 \omega_{pe}}\right)^4M_p^4(\kappa,\bar{k},\bar{\omega})$. Where $r_L$ is the Larmor radius. Because the previous expression and the fact that the induced magnetization increases with the decreasing of the spectral index it is expected that the non-thermal effect enhances the total energy radiated in unmagnetized Lorentzian plasmas.


\section{Conclusion}

We have demonstrated that for non-relativistic thermal velocities the effect of the Kappa distribution for transverse waves is negligible and has a maximum at the inertial length. This characteristic impacts on the spatial term of the ponderomotive force where the non-thermal effect increases with the wavenumber and tends to an asymptotic value. We have also obtained an expresion of the spatial term of the ponderomotive force for low temperature plasmas, which can be applied to model systems which require to consider thermal effects. However, due to what has been said about the dispersion relation the effect of the kappa parameter is not significant giving as a relative difference of the ponderomotive forces for the Kappa and Maxwellian distributions of an order less than $\sim 10^{-4}$ for $\kappa > 2$ and $\alpha_e/c < 10^{-2}$. This results allows us to use without issues the Maxwellian distribution for the interaction of non-thermal, non-relativistic and unmagnetized plasmas with waves with spatial inhomogeneities. Also for further investigation specially related to space physics plasmas this results make us to expect that if we include other parameters that characterize the plasma interaction with the wave propagation like a magnetic field we could obtain a greater effect of the kappa parameter. \\ 

For the temporal factor of the ponderomotive force we have found that because for unmagnetized plasmas it only appears due to thermal effects, the kappa parameter becomes more relevant for both transverse and longitudinal wave propagation. Also this factor is responsible for the generation of a slowly varying magnetic field in the ponderomotive interaction of the electromagnetic waves with the plasma. It has been found that the non-thermal effect enhances the ponderomotive magnetization for electromagnetic waves. From this it follows that because the total radiated power is proportional to the fourth power of the ponderomotive magnetization it should increase as the plasma moves away of the thermal equilibrium. \\

This results would provide useful information for analysis and interpretation of phenomena associated with the ponderomotive force due to the wave propagation in unmagnetized non-thermal plasmas. It would also gives us a useful base from which to extend the analysis to the case of magnetized plasmas. This last it is very relevant in space physics phenomena where we usually find the presence of external magnetic fields, where the spatial and temporal terms of the ponderomotive force are related respectively with the Miller and Abraham forces and terms associated with other forces such as the magnetic moment pumping (MMP) are also added \cite{lundin_ponderomotive_2007}. These forces for Alfv\'en and cyclotron waves appears in phenomena associated with acceleration of ions in the polar regions, auroral density cavities, the penetration of solar wind in the magnetosphere, the electromagnetic ULF waves in the terrestrial magnetosphere among others. Hence, it can be used this formalism extended for external magnetic field to contribute to the study of non-thermal effects in a variety of space physics phenomena, which we will leave for future investigation. \\

In summary our results give us a useful base to include thermal and non-thermal effects in phenomena associated with ponderomotive force for unmagnetized plasmas and to extend it to magnetized plasmas also. Showing a great impact of the non-thermal effect on the temporal factor of the ponderomotive force, which is why it is fundamental to include this results in ponderomotive phenomena in plasmas described by Kappa distributions.  


\bibliographystyle{jpp}

\bibliography{draft}

\end{document}